УДК 004.9:374

**Стрюк Андрій Миколайович**
кандидат педагогічних наук, доцент кафедри моделювання та програмного забезпечення
ДВНЗ «Криворізький національний університет», м.Кривий Ріг, Україна
*andrey.n.stryuk@gmail.com*

**Рассовицька Марина Віталіївна**
асистент кафедри моделювання та програмного забезпечення
ДВНЗ «Криворізький національний університет», м.Кривий Ріг, Україна
*rassovitskayamarina@mail.ru*

## СИСТЕМА ХМАРО ОРІЄНТОВАНИХ ЗАСОБІВ НАВЧАННЯ ЯК ЕЛЕМЕНТ ІНФОРМАЦІЙНОГО ОСВІТНЬО-НАУКОВОГО СЕРЕДОВИЩА ВНЗ

**Анотація.** Метою дослідження є проектування та реалізація хмаро орієнтованого навчального середовища окремого підрозділу ВНЗ. В роботі проведено аналіз існуючих підходів до побудови хмаро орієнтованих навчальних середовищ, формування вимог до хмаро орієнтованих засобів навчання, вибір на підставі цих вимог хмарних ІКТ навчання та експериментальне їх застосування для побудови хмаро орієнтованого навчального середовища окремого підрозділу ВНЗ з використанням відкритого програмного забезпечення та ресурсів власної ІТ-інфраструктури навчального закладу. Результати дослідження планується узагальнити для формування рекомендацій щодо проектування загального хмаро орієнтованого середовища ВНЗ.

**Ключові слова:** хмарні технології; ІКТ навчання; інформаційне освітньо-наукове середовище; хмаро орієнтоване середовище.

## 1. ВСТУП

Невід'ємною рисою сучасного суспільства є прагнення мобільності, що є сучасною міждисциплінарною парадигмою в соціальних і гуманітарних науках, яка досліджує переміщення людей, ідей і речей, а також наслідки цих рухів. Як зазначає М. Шеллер [4], у соціально-гуманітарних науках за останні десять років сформувався новий підхід до вивчення мобільності: комплексне дослідження спільного руху людей, об'єктів та інформації. В інформаційному суспільстві комп'ютерні мережі та інші засоби інформаційно-комунікаційних технологій (ІКТ) сприяють глобалізації, розвитку міжнародного ринка праці, зростанню різних видів індивідуальної мобільності особистості. Одним з передових напрямків розвитку мобільних ІКТ є повсюдне використання технології хмарних обчислень.

У даній роботі проаналізовано та узагальнено передові дослідження з проблеми використання хмарних ІКТ у навчальній, науковій та організаційній діяльності ВНЗ, спроектовано хмаро орієнтоване навчальне середовища окремого підрозділу ВНЗ з використанням відкритого програмного забезпечення та ресурсів власної ІТ-інфраструктури навчального закладу, розглянуто результати експериментального впровадження спроектованого хмаро орієнтованого середовища окремого підрозділу та його інтеграції в інформаційне освітньо-наукове середовище ВНЗ. Розглянуто вплив хмарних технологій на формування та розвиток освітнього середовища ВНЗ.

**Постановка проблеми.** За останні роки хмарні технології набули значного поширення в усіх сферах людської діяльності. Рушійними силами розвитку сучасного суспільства є наука та освіта, тому особливу увагу слід приділити тенденціям упровадження хмарних технологій саме в цих галузях людської діяльності. Повсюдне використання хмарних ІКТ у сучасних університетах сприяє розвитку інформаційного освітньо-наукового середовища ВНЗ. У той же час, згідно останнім дослідженням [9], використання хмарних технологій не завжди є системним та педагогічно доцільним. У зв'язку із цим виникає протиріччя між інтенсивним поширенням хмаро орієнтованих засобів навчання та відсутністю критеріїв їх системного використання, що зумовлює необхідність детального розгляду ролі хмарних ІКТ у формуванні освітньо-наукового середовища та визначення критеріїв педагогічної доцільності їх застосування.

**Аналіз останніх досліджень і публікацій.** Аналіз проведених на сьогодні досліджень вказує на те, що найбільша ефективність від упровадження хмарних технологій навчання досягається при комплексному їх застосуванні на рівні ВНЗ або на міжвузівському рівні [6; 9]. Впливу хмарних технологій на вищу освіту присвячено детальні дослідження об'єднання EDUCASE [3] та колективу авторів під керівництвом З. С. Сейдаметової [8]. В той же час, у наукових дослідженнях мало уваги приділено застосуванню хмарних технологій для підтримки навчальної, наукової та організаційної діяльності окремого структурного підрозділу ВНЗ, досвід якого можна було б узагальнити та поширити на весь навчальний заклад та на систему освіти в цілому.

**Мета статті.** Метою нашого дослідження є аналіз, узагальнення та систематизація досвіду використання хмарних ІКТ у навчальній, науковій та організаційній діяльності ВНЗ, визначення вимог до системи хмаро орієнтованих засобів навчання як елементу інформаційного освітньо-наукового середовища ВНЗ.

## 2. МЕТОДИ ДОСЛІДЖЕННЯ

Дослідження проводилось спільною науково-дослідною лабораторією з питань використання хмарних технологій в освіті ДВНЗ «Криворізький національний університет» та Інституту інформаційних технологій та засобів навчання НАПН України в рамках НДР «Система психологічно-педагогічних вимог до засобів ІКТ навчального призначення» (ДР № 0112U000281). Метою дослідження є розробка та експериментальна перевірка методики використання хмаро орієнтованого освітньо-наукового середовища в процесі навчання інформатичних дисциплін. Експериментальною базою виступав Державний вищий навчальний заклад «Криворізький національний університет». Під час дослідження використовувались такі методи: аналіз теоретичних джерел з проблем застосування хмарних технологій у вищій освіті, вивчення й узагальнення передового досвіду викладання інформатичних дисциплін, співбесіди та опитування студентів, викладачів, аналіз результатів експериментального впровадження хмаро орієнтованих засобів навчання у окремому структурному підрозділі ВНЗ.

## 3. РЕЗУЛЬТАТИ ДОСЛІДЖЕННЯ

Теоретичною основою хмарних технологій є концепція «комунальних обчислень» (utility computing) Дж. Маккарті (запропонована у 1961 р., опублікована у 1999 р.), сутність якої полягає у розгляді комп'ютерних ресурсів як вимірюваних та гнучко дозованих послуг на зразок тих, що надають оператори мобільного зв'язку [2]. Технологічна реалізація цієї концепції вимагала знаходження балансу між традиційною термінальною (клієнт-серверною) ідеологією та суто розподіленими системами з автобалансуванням потужності. Майже 40 років еволюції цієї концепції у бізнес-моделях IBM та інших постачальників мейнфреймів привели до появи центрів опрацювання даних (ЦОД) у їх сучасному вигляді.

Точкою біфуркації стало утворення певного дисбалансу між великою кількістю персональних портативних комп'ютерних пристроїв низької потужності і надлишковою обчислювальною потужністю спеціалізованих центрів обробки даних, що знаходяться у корпоративній власності. Портативні пристрої задовольняли потребу користувачів у повсякчасному та повсюдному доступі до необхідних даних та програм, але апаратна та програмна несумісність окремих пристроїв, їх низька потужність, відсутність єдиних вимог до інтерфейсу, нерозвиненість механізмів обміну даних між окремими пристроями та доступу до різних сховищ даних зменшували ефективність використання цих засобів. У той же час розвиток технологій розподілених обчислень та світовий досвід використання розподілених систем технологічно уможливлює доступ з окремих персональних пристроїв до потужних обчислювальних ресурсів та значних за об'ємом сховищ даних, що знаходились у розпорядженні корпоративних центрів обробки даних. Так з'явився новий вид utility computing – сфера комп'ютерних послуг спільного використання, які тепер прийнято називати «хмарними» (cloud computing). У 2009 році Л. Вакуеро разом із співавторами на основі аналізу більш ніж двадцяти різних трактувань поняття «хмара» в контексті ІКТ дійшов висновку, що в загальному значені «хмара» – це великий масив легкодоступних віртуальних ресурсів (апаратних, програмних платформ та послуг). Ці ресурси можуть динамічно змінюватись, щоб пристосуватися до змін навантаження (масштабування), що зумовлює оптимальне їх використання [5].

Поширення хмарних ІКТ породжує нові – хмаро орієнтовані – технології навчання, що пропонують систему нових засобів, оновлених методів і форм організації навчання та управління навчальною діяльністю. Сьогодні найбільший вплив хмарні ІКТ здійснюють саме на засоби навчання: значна кількість методів та форм організації навчання, що сформувалися в процесі розвитку комп'ютерно-орієнтованих технологій навчання, не набула суттєвих змін. Водночас зазначимо, що у сфері уваги сучасної зарубіжної когнітивної психології знаходяться технології навчання, в яких на перше місце виходить загальноінформаційна діяльність із здобуття відомостей за умови їх постійної та повсюдної доступності та формування навичок неформального навчання.

*Хмаро орієнтовані ІКТ навчання* визначимо як сукупність методів, засобів і прийомів діяльності, що використовуються для організації і супроводу навчального процесу, збирання, систематизації, зберігання, опрацювання, передавання, подання повідомлень і даних навчального призначення та використовують динамічний масив віртуалізованих апаратних і програмних ресурсів, доступних через мережу незалежно від термінального пристрою.

Суттєвою відмінністю хмаро орієнтованих ІКТ від хмарних ІКТ є можливість автономної роботи термінальних засобів (як спеціально виготовлених, так й пристосованих), що дозволяє у їх якості використовувати усі наявні засоби ІКТ-інфраструктури вітчизняних ВНЗ. У дослідженні В. Ю. Бикова [2] окреслено функції ІКТ-підрозділів, що підтримують і розвивають ІКТ-системи на базі адаптивних інформаційно-комунікаційних мереж, тобто тих, які у своїй роботі спираються на хмарну (корпоративну або загальнодоступну) ІКТ-інфраструктуру, і визначено окремі дидактичні функції, а також деякі принципово важливі функції здійснення наукових досліджень, що передбачають доцільне координоване та інтегроване використання сервісів і технологій хмарних обчислень. Якщо стратегія розвитку інформатизації освітніх організаційних структур, зазначає дослідник, передбачає використання хмарних технологій, можливі такі сервісні моделі реалізації хмарного підходу:

– створення і підтримання власної корпоративної хмари, що обов'язково включає побудову, підтримання функціонування і забезпечення розвитку власного центру опрацювання даних, його програмно-апаратних засобів й електронних інформаційних ресурсів, а також передбачає існування потужного ІКТ-підрозділу;

– орієнтація на загальнодоступну хмару, що передбачає використання на умовах повного аутсорсингу засобів і сервісів зовнішньої відносно освітньої структури розподіленої мережі ЦОД, а також наявність у структурі ІКТ-підрозділу. Функції цього ІКТ-підрозділу суттєво відрізняються від тих, які виконує (має виконувати) ІКТ-підрозділ, що спирається на корпоративну сервісну модель, а чисельність ІКТ-персоналу і вимоги до їхньої кваліфікації є порівняно меншими, ніж у разі використання корпоративної хмари;

– орієнтація на гібридну (комбіновану) модель реалізації ІКТ-сервісів (одночасне використання як корпоративної, так і загальнодоступної хмари) [6].

Окрім необхідності створення та супроводження власного центру опрацювання даних, його програмно-апаратних засобів й електронних інформаційних ресурсів, проектування корпоративного хмаро орієнтованого середовища для ВНЗ значно ускладнюється у зв'язку з розгалуженістю цілей застосування хмарних ІКТ. У результаті дослідження багатьох науковців були спрямовані перш за все на використання хмарних ІКТ при вивченні окремих навчальних курсів та циклів дисциплін. Наступним кроком до комплексного застосування хмарних ІКТ у ВНЗ ми вважаємо проектування хмаро орієнтованого середовища окремого підрозділу ВНЗ.

Проектування такого середовища потребує уточнення цілей використання та формування вимог. Ці вимоги ми визначаємо за видами діяльності, що виконує такий підрозділ ВНЗ, як кафедра: наукова діяльність, організаційна діяльність та навчання. Щоб забезпечити ці види діяльності, на базі хмаро орієнтованих ІКТ має бути реалізовано комунікаційне середовище, персональне сховище даних, загальне сховище, сховище навчальних матеріалів та науково-дослідницьких проектів.

Зорієнтованість хмаро орієнтованих ІКТ навчання на повсюдний та відкритий доступ розширює можливості співпраці суб'єктів навчального процесу, зокрема, в спільному плануванні та реалізації різних видів навчальної діяльності і спільній розробці та тестування програмного забезпечення, організації комп'ютерного експерименту, що є невід'ємною складовою навчального процесу з інформатичних дисциплін. Підхід, за якого хмарні ІКТ розглядаються як комунікаційне середовище суб'єктів навчального процесу та гнучкий засіб організації сховища даних навчального призначення, веде до формування хмаро орієнтованого освітньо-наукового середовища вищого навчального закладу, в якому окремі дидактичні функції передбачають доцільне координоване та інтегроване використання сервісів і технологій хмарних обчислень [6]. Аналіз існуючих хмаро орієнтованих засобів навчання та підходів до їх використання надав можливість виділити компоненти системи хмаро орієнтованих засобів навчання у освітньому середовищі ВНЗ (рис. 1).

Для експериментальної перевірки ефективності роботи такої системи засобів на кафедрі моделювання та програмного забезпечення ДВНЗ «Криворізький національний університет» були використані наступні інструменти:

– система управління навчанням, що реалізована на базі відкритої платформи MOODLE;

– соціальні мережі, серед яких за результатами опитування серед студентів все більшої популярності набувають Facebook-подібні мережі;

– wiki-система, реалізована на базі відкритої платформи MediaWiki;

– інтегроване хмарне середовище на базі відкритої системи OwnCloud.

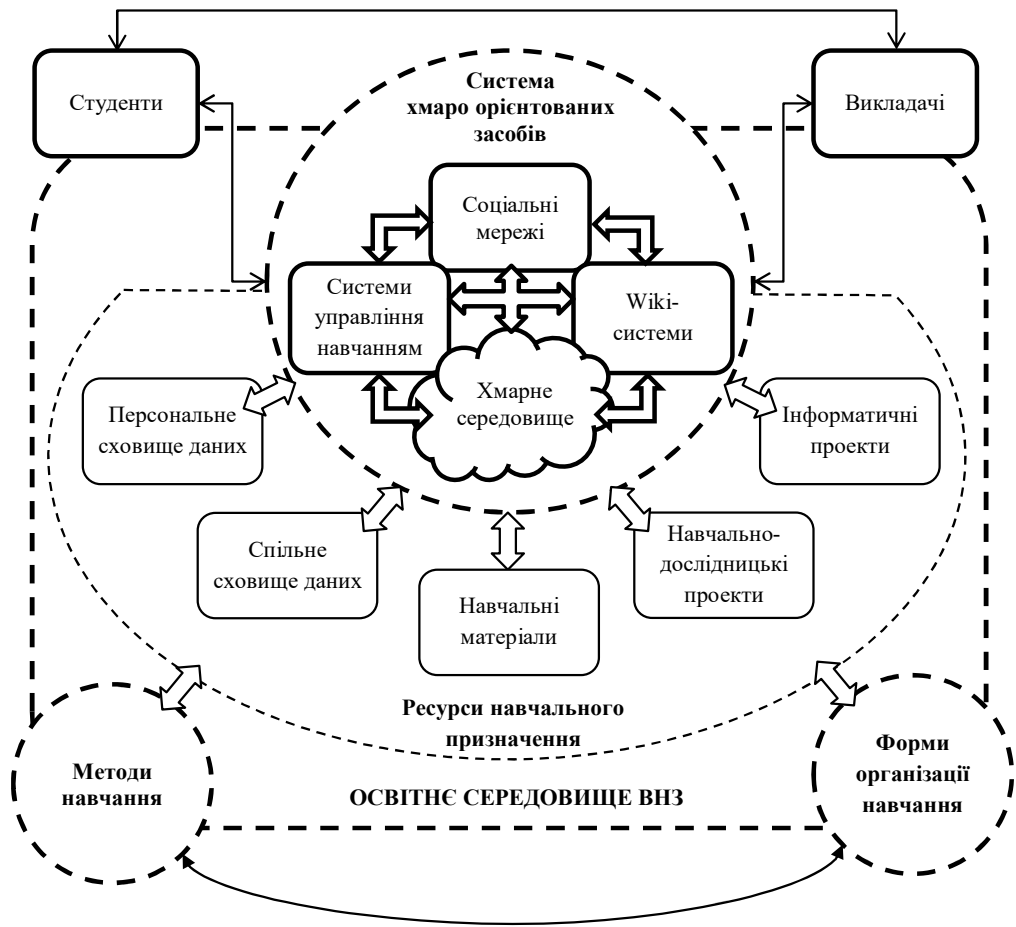

*Рис. 1. Система хмаро орієнтованих засобів навчання як елемент освітнього середовища ВНЗ*

Хмарне середовище виконує інтегруючу та системоутворюючу функцію. З одного боку, за допомогою хмарного середовища здійснюється ресурсна підтримка інших засобів ІКТ навчання, з іншого, хмарне середовище виступає як самостійний засіб навчання, за допомогою якого вирішуються окремі навчальні задачі.

Незважаючи на те, що популярні загальнодоступні корпоративні хмари загального призначення надають достатній для окремого підрозділу ВНЗ обсяг хмарних послуг, було прийнято рішення про створення приватної кафедральної хмари. З одного боку, використання публічних хмар на рівні ВНЗ вважається ризикованим через те, що данні та обчислювальні ресурси контролюватимуться третьою стороною, а також через те, що постачальник послуг може в односторонньому порядку призупинити або обмежити послуги, що надаються навчальним закладам. З іншого, оцінка ІТ-інфраструктури ДВНЗ «Криворізький національний університет» показала, що обсяг власних обчислювальних ресурсів та кваліфікація штатних спеціалістів цілком достатні для розгортання приватної кафедральної хмари.

Платформа OwnCloud [1], використана для побудови приватної кафедральної хмари, має наступні технічні переваги:
– простота розгортання та адміністрування;
– помірні системні вимоги;
– відкритий код;
– підтримка спільнотою розробників.

В. А. Коваленко [7] також виділяє наступні переваги використання системи OwnCloud: міжплатформність, інтегрований перегляд документів, календар і планувальник, редактор текстів з підтримкою синтаксису найбільш популярних мов програмування, спільний доступ, захищеність даних, контроль версій та підтримку розробки додатків.

Доступ до хмари, побудованій на платформі OwnCloud, здійснюється за допомогою веб-браузеру або спеціальної програми, що встановлюється на хмаро орієнтований термінальний пристрій – персональний або портативний комп'ютер (рис. 2). Середовище надає можливість спільно використовувати окремі файли, планувальник, контакти та інші додатки, необхідні в організаційній, науковій та навчальній діяльності підрозділу ВНЗ.

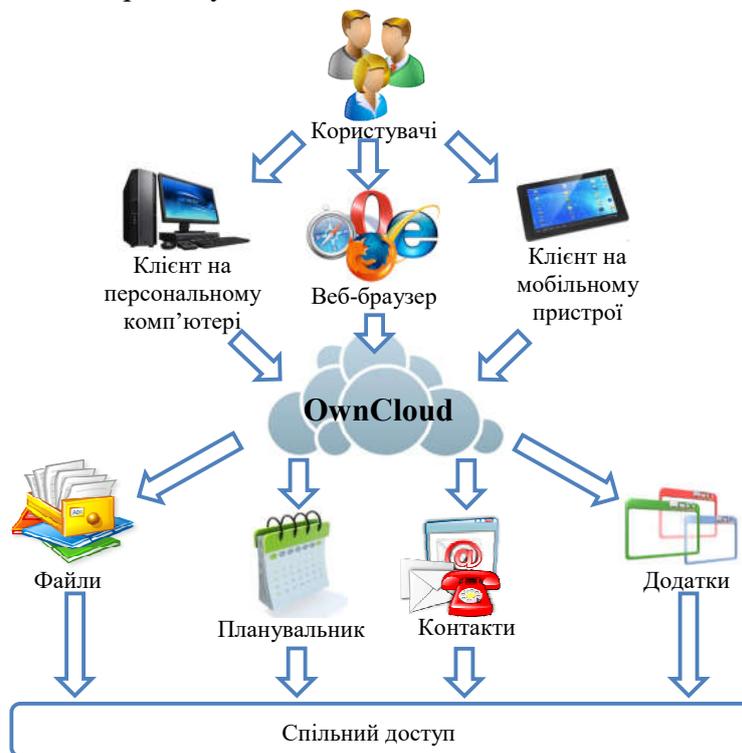

*Рис. 2. Модель використання системи OwnCloud*

Основними цілями використання корпоративної кафедральної хмари є:
– спрощення доступу викладачів до кафедральних документів;
– забезпечення спільної роботи викладачів над методичними посібниками, підручниками тощо;
– організація спільної роботи студентів з курсового та дипломного проектування, виконання спільних проектів, передбачених різними дисциплінами.

Крім того, платформа OwnCloud підтримує створення нових додатків, що є актуальним для підготовки фахівців з програмної інженерії. Вбудований текстовий редактор системи OwnCloud розпізнає близько 30 мов програмування, що надає можливість ефективно використовувати його для спільної роботи над інформатичними проектами наряду з такими спеціалізованими веб-ресурсами, як Scratch.mit.edu, IDEOne.com, CodePad.org, CollabEdit.com, Pythonv3, TouchDevelop та ін. Узагальнена модель взаємодії викладачів і студентів у хмарному середовищі показана на рис. 3.

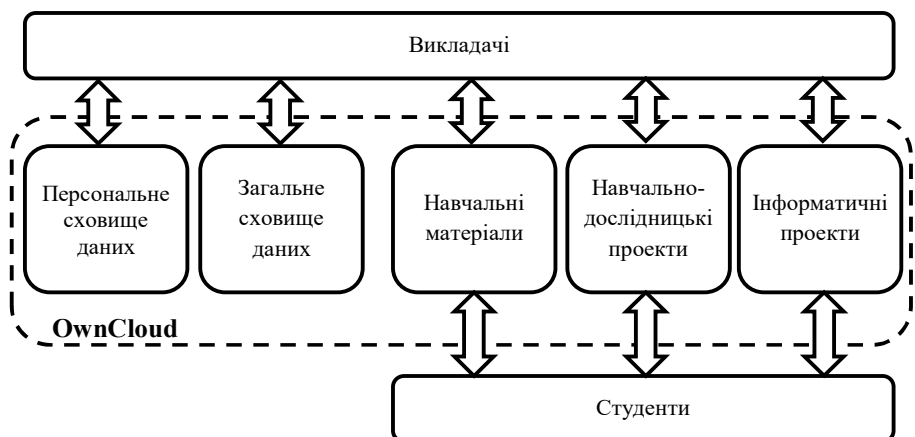

*Рис. 3. Узагальнена модель взаємодії викладачів і студентів у хмарному середовищі*

Запропонована модель впроваджена на кафедрі моделювання та програмного забезпечення ДВНЗ «Криворізький національний університет» та використовується для студентів напрямку програмна інженерія при вивченні таких дисциплін, як «Теорія інформації та кодування», «Операційні системи», «Інженерія програмного забезпечення паралельних і розподілених систем», а також для студентів різних інженерних спеціальностей при викладанні дисциплін «Інформатика», «Обчислювальна техніка та програмування». Проблемам впровадження хмарних технологій у наукові дослідження присвячено спеціальний курс, що викладається для аспірантів першого року навчання. Слід зазначити, що хмарні технології постійно розвиваються, щоденно з'являються нові сервіси, існуючі знаходять нові застосування. У зв'язку з цим дослідження впливу хмарних технологій на формування та розвиток інформаційного освітньо-наукового простору ВНЗ тривають, формуються нові наукові гіпотези щодо впливу хмарних ІКТ на наукову та освітню діяльність, які потребують ретельної перевірки. Але безсумнівним є те, що хмарні технології створюють міцне підґрунтя для подальшого сталого розвитку інформаційного суспільства та формування майбутнього суспільства знань.

## 4. ВИСНОВКИ ТА ПЕРСПЕКТИВИ ПОДАЛЬШИХ ДОСЛІДЖЕНЬ

Спроектоване та реалізоване хмаро орієнтоване навчальне середовище окремого підрозділу ВНЗ потребує оцінки ефективності використання у навчальному процесі, яку має забезпечити запланований педагогічний експеримент. Наразі викладачами кафедри проводяться науково-методичні дослідження з використання хмарних технологій у професійній підготовці майбутніх програмістів. На різних етапах цих досліджень планується уточнити етапи проектування хмаро орієнтованого середовища підрозділу ВНЗ і згодом узагальнити його до хмаро орієнтованого середовища ВНЗ.

## СПИСОК ВИКОРИСТАНИХ ДЖЕРЕЛ

# СИСТЕМА ОБЛАЧНО ОРИЕНТИРОВАННЫХ СРЕДСТВ ОБУЧЕНИЯ КАК ЭЛЕМЕНТ ИНФОРМАЦИОННОЙ ОБРАЗОВАТЕЛЬНО-НАУЧНОЙ СРЕДЫ ВУЗА

**Стрюк Андрей Николаевич**
кандидат педагогических наук, доцент кафедры моделирования и программного обеспечения
ГВУЗ «Криворожский национальный университет, г. Кривой Рог, Украина
*andrey.n.stryuk@gmail.com*

**Рассовицкая Марина Витальевна**
ассистент кафедры моделирования и программного обеспечения
ГВУЗ «Криворожский национальный университет, г. Кривой Рог, Украина
*rassovitskayamarina@mail.ru*

**Аннотация.** Целью исследования является проектирование и реализация облачно ориентированной учебной среды отдельного подразделения вуза. В работе проведен анализ существующих подходов к построению облачно ориентированных учебных сред, формирование требований к облачно ориентированным средствам обучения, выбор на основании этих требований облачных ИКТ обучения и экспериментальное их применения для построения облачно ориентированного учебной среды отдельного подразделения вуза с использованием открытого программного обеспечения и ресурсов собственной ИТ-инфраструктуры учебного заведения. Результаты исследования планируется обобщить для формирования рекомендаций по проектированию общей облачно ориентированной среды вуза.

**Ключевые слова:** облачные технологии; ИКТ обучения; информационная образовательно-научная среда; облачно ориентированная среда.

# THE SYSTEM OF CLOUD ORIENTED LEARNING TOOLS AS AN ELEMENT OF EDUCATIONAL AND SCIENTIFIC ENVIRONMENT OF HIGH SCHOOL

**Andrii M. Striuk**
PhD, associate professor of modeling and software department
State institution of higher education «Kryvyi Rih National University», Kryvyi Rih, Ukraine
*andrey.n.stryuk@gmail.com*

**Maryna V. Rassovytska**
assistant of modeling and software department
State institution of higher education «Kryvyi Rih National University», Kryvyi Rih, Ukraine
*rassovitskayamarina@mail.ru*

**Abstract.** The aim of this research is to design and implementation of cloud based learning environment for separate division of the university. The analysis of existing approaches to the construction of cloud based learning environments, the formation of requirements cloud based learning tools, the selection on the basis of these requirements, cloud ICT training and pilot their use for building cloud based learning environment for separate division of the university with the use of open source software and resources its own IT infrastructure of the institution. Results of the study is planned to generalize to develop recommendations for the design of cloud based environment of high school.

**Keywords:** cloud technology; information and communication technology of education; information education and research environment; cloud-based environment.